\documentclass[prb,twocolumn, groupedaddress, showkeys,showpacs]{revtex4}

\usepackage{graphicx}
\begin{document}


\title{Directed transport driven by the transverse wall vibration}


\author{Bao-quan  Ai}

\affiliation{Laboratory of Quantum Information Technology, ICMP and
 SPTE, South China Normal University, 510631 Guangzhou, China.}


\date{\today}
\begin{abstract}
 \indent Directed transport of overdamped
Brownian particles in an asymmetrically periodic tube is
investigated in the presence of the tube wall vibration. From the
Brownian dynamics simulations we can find that the perpendicular
wall vibration can induce a net current in the longitudinal
direction when the tube is asymmetric. The direction of the current
at low frequency is opposite to that at high frequency. One can
change the direction of the current by suitably tailoring the
frequency of the wall vibration.
\end{abstract}




\maketitle

\section {Introduction}
\indent  Stochastic transport on periodic asymmetric substrates far
from equilibrium has raised widespread interest in the recent
literature \cite{a1}. In these spatially periodic structures,
directed motion of particles can induced by zero-mean nonequilibrium
fluctuations and noise. These come from the desire to understand
molecular motors\cite{a2}, nanoscale friction\cite{a3}, surface
smoothing\cite{a4}, coupled Josephson junctions\cite{a5}, optical
ratchets and directed motion of laser-cooled atoms\cite{a6}, and
mass separation and trapping schemes at the microscale\cite{a7}. The
fundamental condition for directed transport to occur is that either
the spatial reflection symmetry of the system is broken or
fluctuations are statistically asymmetric. Several models have been
proposed to explain this transport mechanism under various
nonequlibrium situations. Typical examples are rocking ratchets
\cite{a8}, flashing ratchets\cite{a9}, diffusion ratchets\cite{a10},
correlation ratchets\cite{a11}.

\indent Most studies have revolved around the energy barrier. The
nature of the barrier depends on which thermodynamic potential
varies when passing from one well to the other, and its presence
plays an important role in the dynamics of the system. However, in
many transport phenomena\cite{a12}, such as those taking place in
micro- and nano-pores, zeolites, biological cells, ion channels,
nanoporous materials and microfluidic devices etched with grooves
and chambers, Brownian particles, instead of diffusing freely in the
host liquid phase, undergo a constrained motion. In these cases the
entropic barriers may appear when coarsening the description of a
complex system in order to simplify its dynamics. Reguera and
coworkers\cite{a13} used the mesoscopic nonequilibrium
thermodynamics theory to derive the general kinetic equation of the
motor system and analyzed in detail the case of diffusion in a
domain of irregular geometry in which the presence of the boundaries
induces an entropy barrier when approaching the dynamics by a
coarsening of the description. In their recent work \cite{a14} they
studied the current and the diffusion of a Brownian particle moving
in a symmetric channel with a biased external force. In our previous
work\cite{a15}, we studied the transport of overdamped particles in
a three dimensional tube in the presence of unbiased external forces
and the motion of particles can indeed be rectified, with a sign
that depends on the details of the wall profile. In the present
work, we study the directed transport of overdamped Brownian
particles moving in asymmetric tube in the presence of the tube wall
vibration. We emphasized on finding whether the transverse wall
vibration can induce a longitudinal net current and how the
parameters of the system affect the transport.
\section{Model and Methods}
\begin{figure}[htbp]
  \begin{center}\includegraphics[width=8cm,height=4cm]{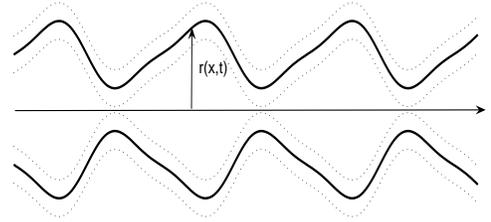}
  \caption{Schematic diagram of a tube with periodicity $L$. The vibrating wall is described by its time dependent half width $r(x,t)=a[\sin(\frac{2\pi x}{L})+\frac{\Delta}{4}\sin(\frac{4\pi
    x}{L})]+b+b_{0}\sin(\omega t)$. $\Delta$ is the asymmetric parameter of the tube shape and $\Delta=-1.0$ in the diagram. The solid line describes the original shape of the tube and the dotted line shows its vibrating shape.  }\label{1}
\end{center}
\end{figure}

\indent In this study, we study the transport of the particles in an
asymmetrically periodic tube driven by the tube wall vibration [Fig.
1]. Since most of the molecular transport occurs in the overdamped
regime, we can safely neglect inertial effects. The overdamped
dynamics can be described by the following Langevin equations
written in a dimensionless form \cite{a14,a15},
\begin{equation}\label{}
    \eta\frac{dx}{dt}=\sqrt{\eta
    k_{B}T}\xi_{x}(t),
\end{equation}
\begin{equation}\label{}
    \eta\frac{dy}{dt}=\sqrt{\eta
    k_{B}T}\xi_{y}(t),
\end{equation}
where $x$, $y$, are the two-dimensional (2D) coordinates, $\eta$ is
the friction coefficient of the particles, $k_{B}$ is the Boltzmann
constant, $T$ is the absolute temperature and $\xi_{x,y}(t)$ is the
Gaussian white noise with zero mean and correlation function:
$<\xi_{i}(t)\xi_{j}(t^{'})>=2\delta_{i,j}\delta(t-t^{'})$ for
$i,j=x, y$. $<...>$ denotes an ensemble average over the
distribution of noise. $\delta(t)$ is the Dirac delta function.
Imposing reflecting boundary conditions in the transverse direction
ensures the confinement of the dynamics within the tube, while
periodic boundary conditions are enforced along the longitudinal
direction for the reasons noted above. The vibrating wall is
described by its time dependent half width
\begin{equation}\label{}
    r(x)=a[\sin(\frac{2\pi x}{L})+\frac{\Delta}{4}\sin(\frac{4\pi
    x}{L})]+b+b_{0}\sin(\omega t),
\end{equation}
where $a$ is the parameter that controls the slope of the tube and
$\Delta$ is the asymmetric parameter of the tube shape.  $b$ is the
parameter that determines the half width at the bottleneck. $b_{0}$
and $\omega$ are the amplitude and frequency of the vibration,
respectively.

\indent  In this case, no general valid analytical expressions are
possible.  However, the study of these transport phenomena is in
many respects equivalent to an investigation of geometrically
constrained Brownian dynamics\cite{b1}. The behavior of the
quantities of interest have been corroborated by Brownian dynamic
simulations performed by integration of the Langevin equation using
the standard stochastic Euler algorithm. The average particle
velocity in the $x$-direction has been derived from an ensemble
average of about $3 \times 10^{4}$ trajectories according to the
following expression. From Eq. (1-2) and the standard stochastic
Euler algorithm, the single integration steps read \cite{b1}:
\begin{equation}\label{}
    x_{new}=x_{old}+\sqrt{2k_{B}T \Delta
    t}R_{1},
\end{equation}
\begin{equation}\label{}
y_{new}=y_{old}+\sqrt{2k_{B}T \Delta t}R_{2},
\end{equation}
if $|y_{new}|<|r(x_{new})|$, the modification positions of
$x,y$-coordinate are
\begin{equation}\label{}
    x_{old}=x_{new}, \indent y_{old}=y_{new},
\end{equation}
if $|y_{new}|\geq|r(x_{new})|$, the particles will meet the wall. In
this case, we use a simple way to simulate the movement of particles
near the wall: When the particles meet the wall, it will return to
its previous position and the time step is incremented, namely,
\begin{equation}\label{}
    x_{old}=x_{old}, \indent y_{old}=y_{old},
\end{equation}
where $x_{old}$, $y_{old}$ are the original positions and $x_{new}$,
$y_{new}$ are the new positions. $R_{1}$, $R_{2}$ are two Gaussian
distributed random numbers with unit variance. $\Delta t$ is the
integration step time.  The average particle velocity along the
$x$-direction,
\begin{equation}\label{}
\nu=\langle\dot{x}\rangle=\lim_{t\rightarrow\infty}\frac{\langle
x(t)
    \rangle}{t}.
\end{equation}

\begin{figure}[htbp]
  \begin{center}\includegraphics[width=8cm,height=6cm]{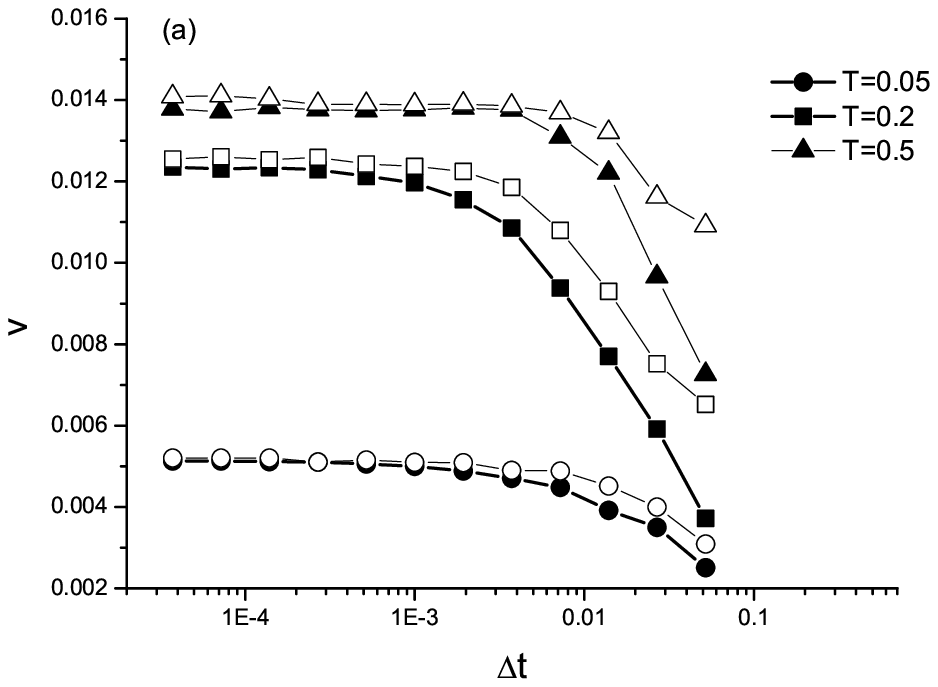}
  \includegraphics[width=8cm,height=6cm]{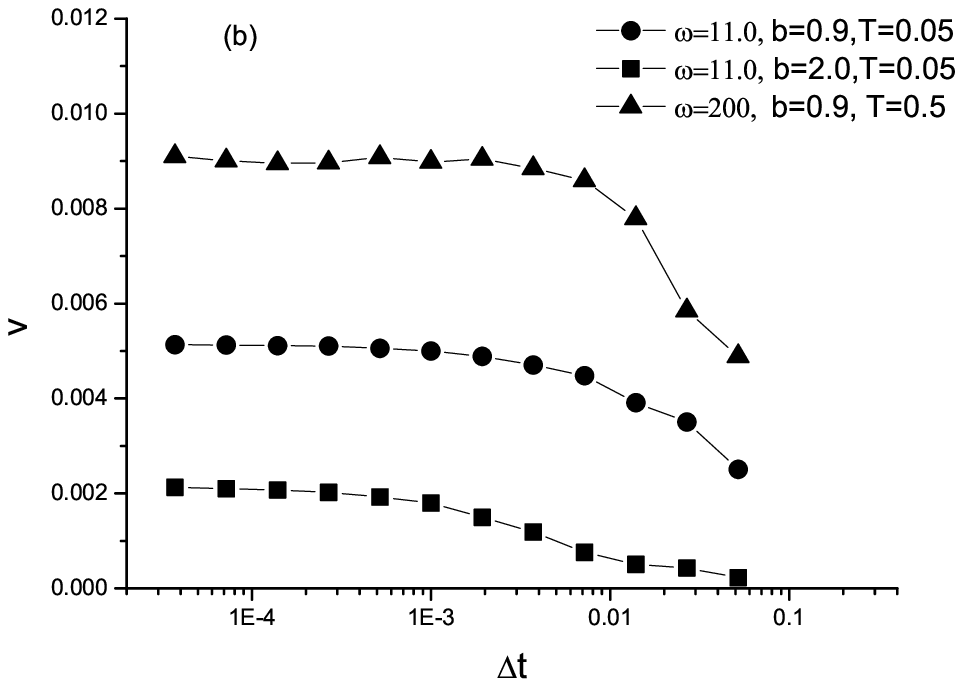}
  \caption{Dependence of the mean velocity $v$ on time step
  $\Delta t$. (a)for different values of temperature at $\Delta=-1.0$, $\omega=11.0$, $a=0.5$, $b=0.9$, $b_{0}=0.3$, and
  $L=1.0$. The full symbols depict the present algorithm and the empty
  symbols indicate the strict reflection algorithm. (b) for different parameters at $\Delta=-1.0$, $a=0.5$, $b_{0}=0.3$, and
  $L=1.0$.}
  \label{1}
\end{center}
\end{figure}

In order to check the validity of the present Brwonian dynamics
algorithm, it is necessary to compare the present results with a
real reflecting boundary condition on the wall. From Fig. 2(a), we
can find that the results from the present algorithm well agree with
that from the strict reflection algorithm for very small time step.
Moreover, the present algorithm takes much less CPU time than the
strict reflection one. We have also checked the convergence of the
algorithm in a large range of the system parameters. From Fig. 2 we
can see that the algorithm is convergent and the numerical results
will not depend on the time step for very small time step.
Therefore, our algorithm will give well approximate results with
respect to the strict reflection algorithm. In order to provide the
requested accuracy of the system dynamics time step was chosen to be
smaller than $10^{-4}$.

\indent We must point that the extension of our scheme to
three-dimension with rotational symmetry along the transport axis is
possible as well. This will consume more computation time, but the
over results remain qualitatively robust.

\section {Numerical results and discussion}
Our emphasis is on finding the asymptotic mean velocity which is
defined as the average of the velocity over the time and thermal
fluctuations. In order to obtain the mean velocity, we carried out
extensive numerical simulations based on Eq. (4-8). For simplicity
we set $\eta=1$ and $k_{B}=1$ throughout the work.
\begin{figure}[htbp]
  \begin{center}\includegraphics[width=8cm,height=8cm]{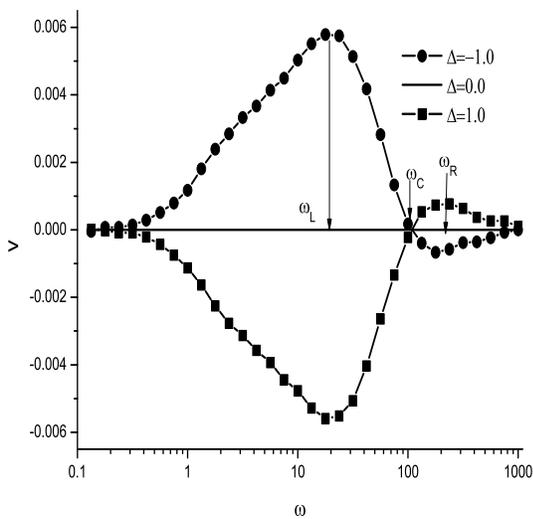}
  \caption{The mean velocity $v$ as a function of vibrating frequency $\omega$ for different values of asymmetric parameter $\Delta=-1.0$, $0.0$, and $1.0$ at
  $T=0.05$, $a=0.5$, $b=0.9$, $b_{0}=0.3$, and $L=1.0$. $\omega_{L}$ is the frequency for the left peak, $\omega_{C}$ is the cross-over frequency and
  $\omega_{R}$ denotes the frequency for the right valley. }.
  \label{1}
\end{center}
\end{figure}

\indent Figure 3 shows the mean velocity $v$ as a function of the
vibrating frequency $\omega$. From the Fig. 3, we can see that for
low frequencies the current is positive for $\Delta<0$, zero at
$\Delta=0$, and negative for $\Delta>0$. The phenomenon will be
opposite to that for high frequencies. It is obvious that the
symmetry between the curve $\Delta=1.0$ and $\Delta=-1.0$ is due to
the symmetry of geometry. Now we will focus on the case of
$\Delta=-1.0$, namely, the right side is steeper than the left one.
In the adiabatic limit $\omega\rightarrow 0$, the vibration can be
expressed by two opposite static driving $-b_{0}$ and $b_{0}$ and
the nonequilibrium system reduces to the system with two equilibrium
states, namely, $v=\frac{1}{2}[v(-b_{0})+v(b_{0})]=0$. For low
frequencies, the particles get enough time to reach the whole area
and the wall will act on the particles. In this case, the
probability of the particles acting on the slanted wall (the left
wall) is larger than that on the steeper wall (the right wall),
resulting in a positive current. On increasing the frequency
$\omega$, due to high frequency, the Brownian particles do not get
enough time to reach the slanted wall (The area of the slanted side
is larger than that of the steep one). Therefore, the probability of
the particles acting on the slanted wall is smaller than that on the
steep wall, yielding a negative current. When the vibration changes
very fast $\omega\rightarrow\infty$, the particles can not
experience the changes of the wall and the system reduces to a
equilibrium system, so the current goes to zero. Therefore, the
direction of the current can be controlled by changing the vibrating
frequencies.

   \indent In order to give more detail insight into the
   characteristic frequencies $ \omega_{L}$, $\omega_{C}$, and
   $\omega_{R}$, we corroborate the numerical estimates of the
   frequencies for a larger range of parameter values. The numerical
   estimates are shown in Fig 4. We find that the characteristic
   frequencies increase with temperature $T$ and the length of the
   period $L$ and decrease with the vibrating amplitude $b_{0}$.  The
   estimated exponential relations are also shown in the figure.
   From the figure we can obtain the approximate scaling results for
   $ \omega_{L}\propto T^{0.64}b_{0}^{-1.57}L^{-0.95}$, $\omega_{C}\propto T^{0.71}b_{0}^{-1.31}L^{-1.32}$, and
   $\omega_{R}\propto T^{0.72}b_{0}^{-1.07}L^{-1.41}$. It is obvious
   that these characteristic frequencies will shift to large values
   for high temperature and to small values for large amplitude of
   the vibration.

\begin{figure}[htbp]
  \begin{center}\includegraphics[width=8cm,height=6cm]{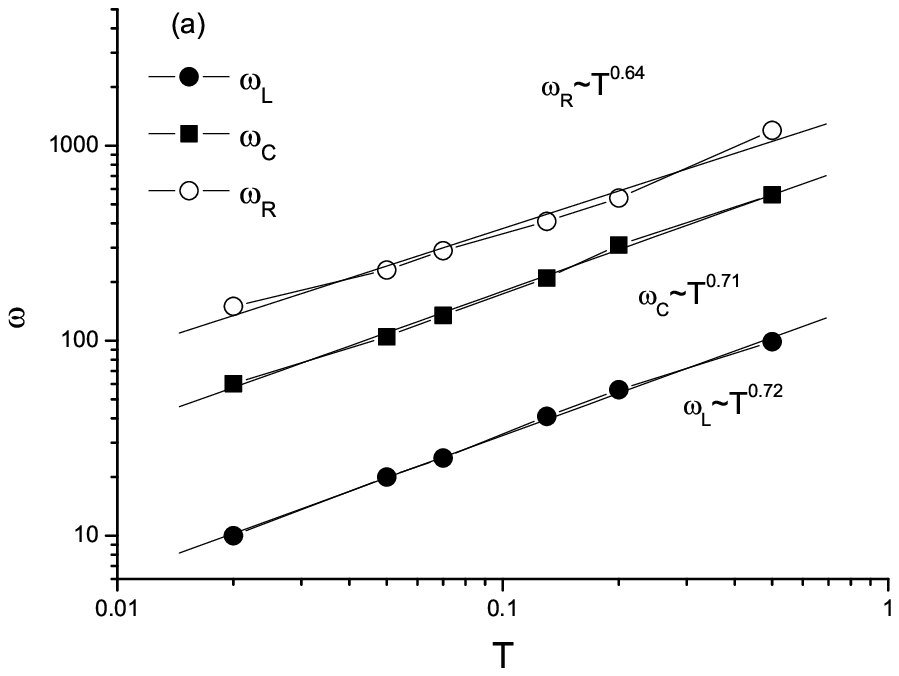}
    \includegraphics[width=8cm,height=6cm]{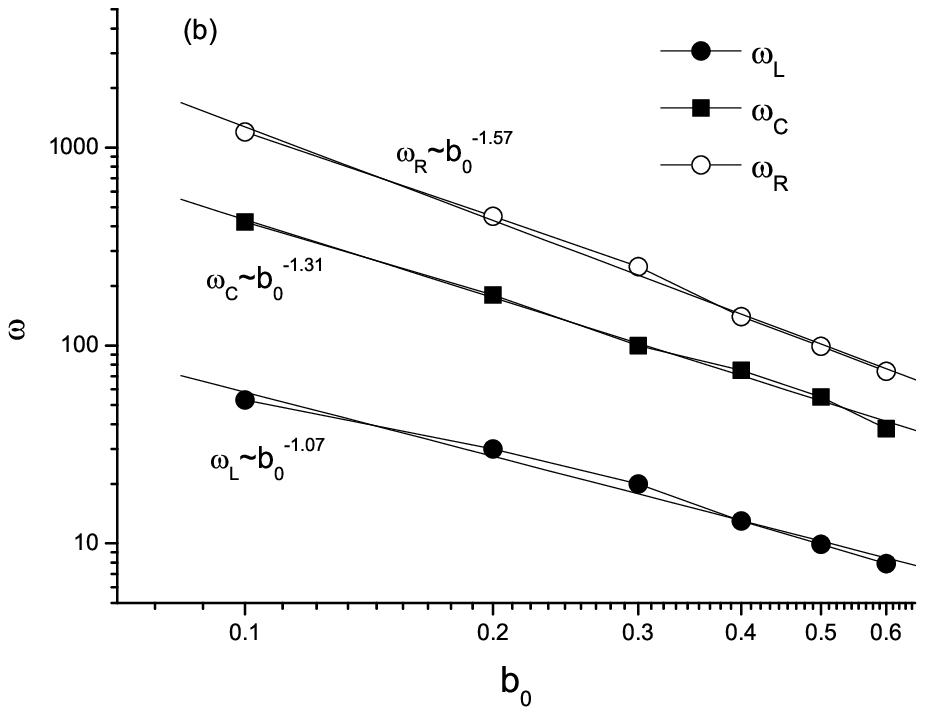}
      \includegraphics[width=8cm,height=6cm]{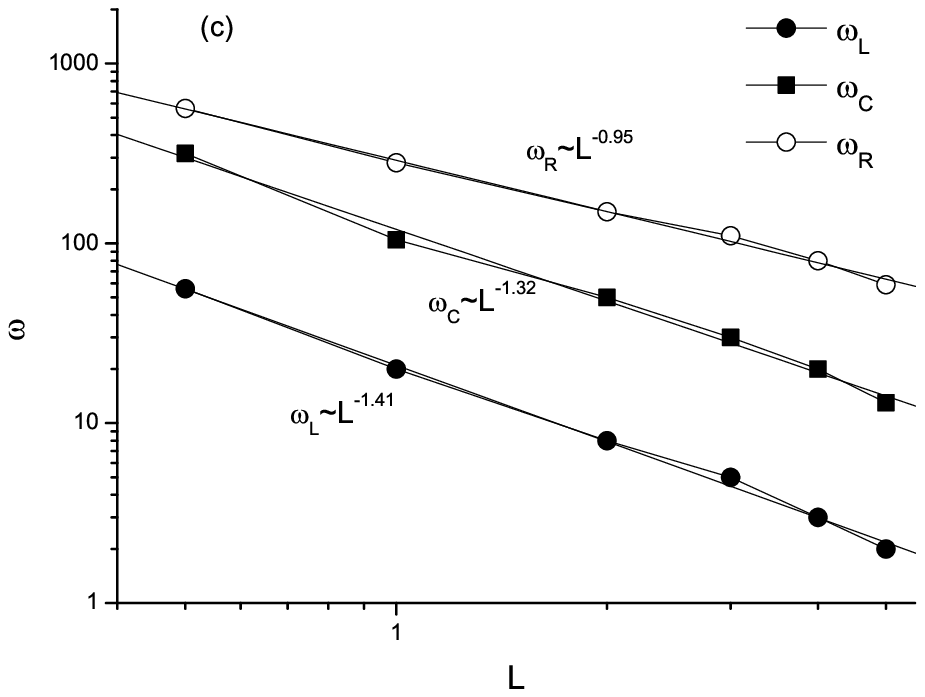}
  \caption{Dependence of characteristic frequencies $ \omega_{L}$, $\omega_{C}$, and
   $\omega_{R}$ on the different parameters. (a)On temperature $T$ at $a=0.5$, $b=0.9$, $b_{0}=0.3$, $\Delta=-1.0$, and $L=1.0$.
   (b)On the vibrating amplitude $b_{0}$ at $T=0.05$, $a=0.5$, $b=0.9$,$\Delta=-1.0$, and $L=1.0$.
   (c)On the period $L$ at $T=0.05$, $a=0.5$, $b=0.9$,$b_{0}=0.3$, and
   $\Delta=-1.0$. The solid lines are the fitting lines.}
  \label{1}
\end{center}
\end{figure}

\begin{figure}[htbp]
  \begin{center}\includegraphics[width=8cm,height=6cm]{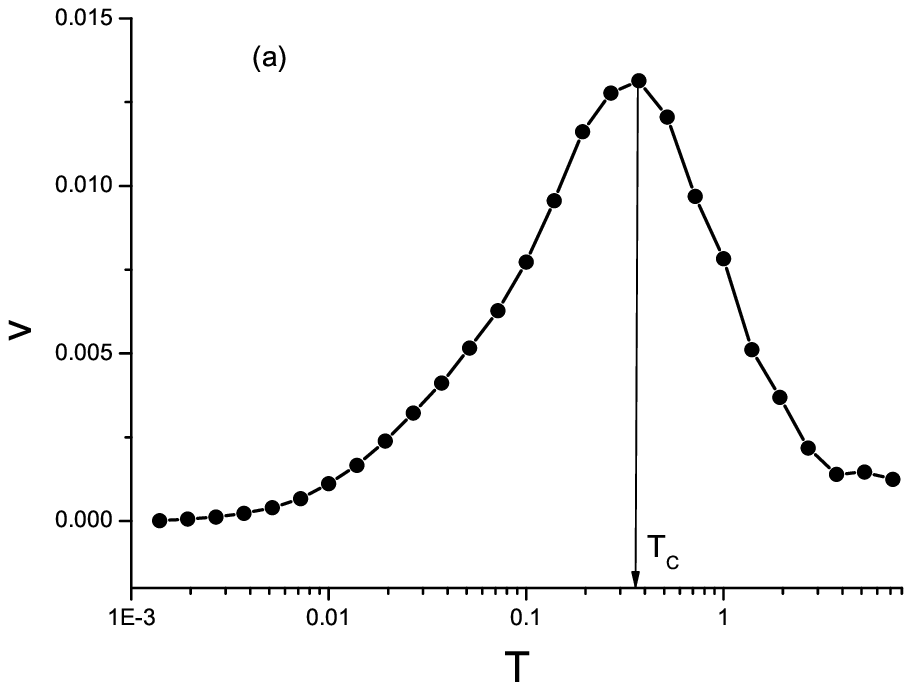}
    \includegraphics[width=8cm,height=6cm]{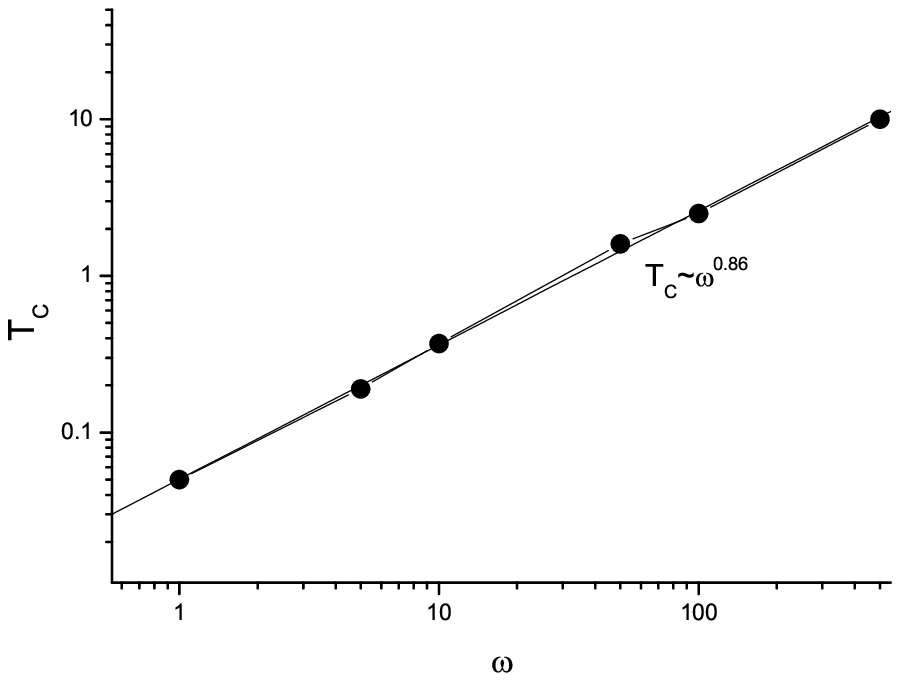}
      \includegraphics[width=8cm,height=6cm]{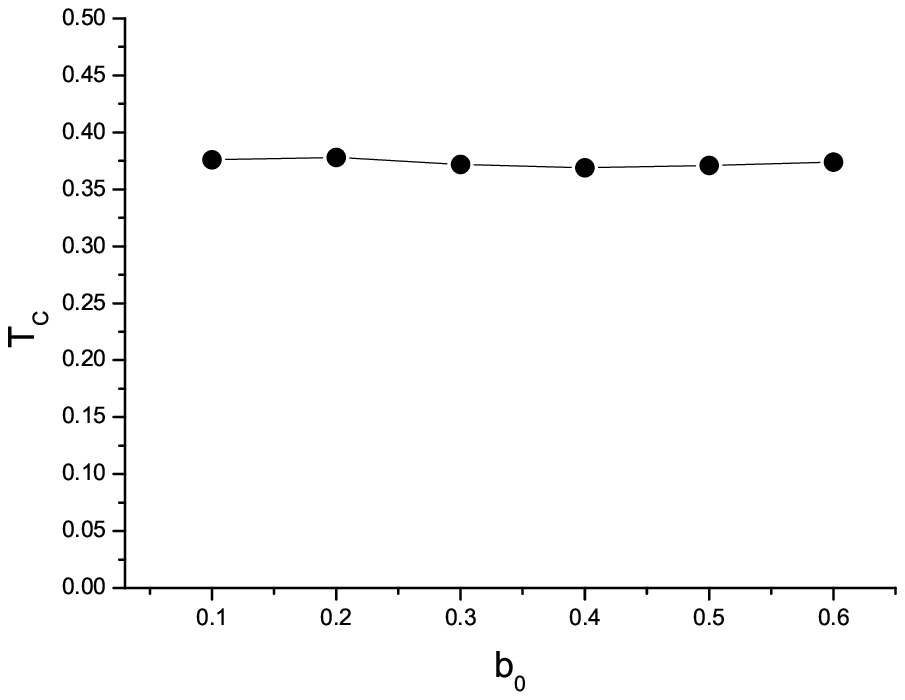}
\caption{(a)The mean velocity $v$ as a function of temperature $T$
at $a=0.5$, $b=0.9$, $b_{0}=0.3$, $\Delta=-1.0$, $\omega=10.0$, and
$L=1.0$. $T_{C}$ is the optimized value of temperature for the peak.
(b)Dependence of $T_{C}$ on the driving frequency $\omega$ at
$a=0.5$, $b=0.9$, $b_{0}=0.3$, and $\Delta=-1.0$. (c)Dependence of
$T_{C}$ on the driving amplitude $b_{0}$ at $a=0.5$, $b=0.9$,
$\omega=10.0$, and $\Delta=-1.0$. The solid lines are the fitting
lines.}
  \label{1}
\end{center}
\end{figure}

\indent  In order to investigate how the temperature affects the
transport, we also show the mean velocity  as a function of
temperature in Fig. 5(a) at $\Delta=-1.0$ and $\omega=10.0$. The
diffusion from the temperature has two roles: jumping from one cell
to another in longitudinal direction and making the particles acting
on the wall in transverse direction. When $T\rightarrow 0$, the
particles cannot reach the wall and the effect of wall disappears
and there is no current. When $T\rightarrow\infty$, the effect of
the wall vibration disappears and the current goes to zero, also.
Therefore, there is an optimized value of $T$ at which the mean
velocity $v$ takes its maximum value, which indicates that the
thermal noise may facilitate the particles transport. In order to
give a feeling of the generality of the optimized value of
temperature, we also study the dependence of $T_{C}$ on the
vibrating frequency and amplitude. From Fig. 5 (b) and (c), we can
obtain the approximate scaling results for the optimized temperature
$T_{C}\propto b_{0}^{0}\omega^{0.86}$. The value of the optimized
temperature will become larger for high driving frequency.
Remarkably, the optimized temperature is independent of the driving
amplitude.

\begin{figure}[htbp]
  \begin{center}\includegraphics[width=8cm,height=6cm]{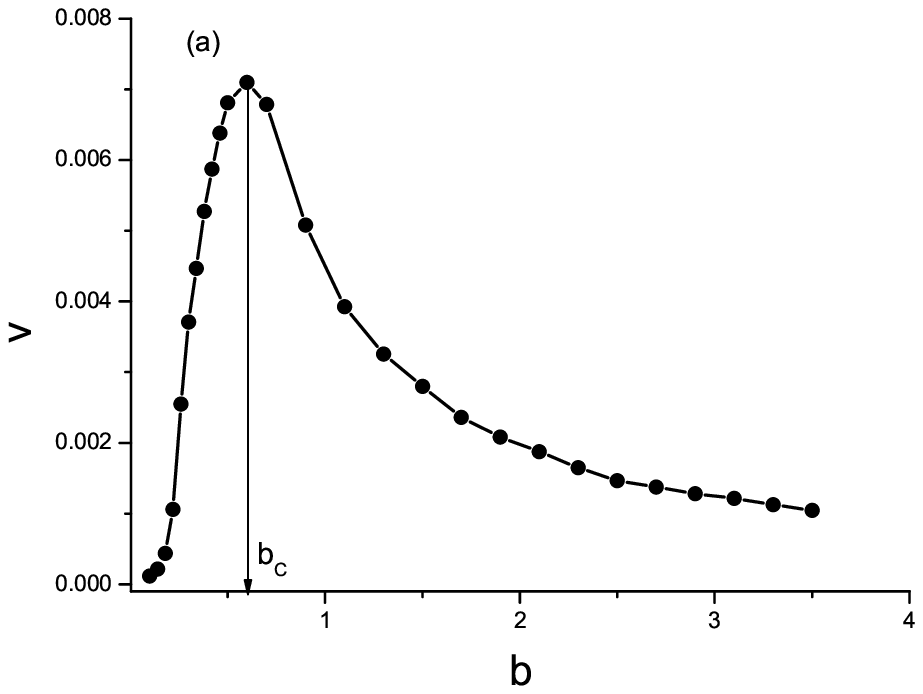}
  \includegraphics[width=8cm,height=6cm]{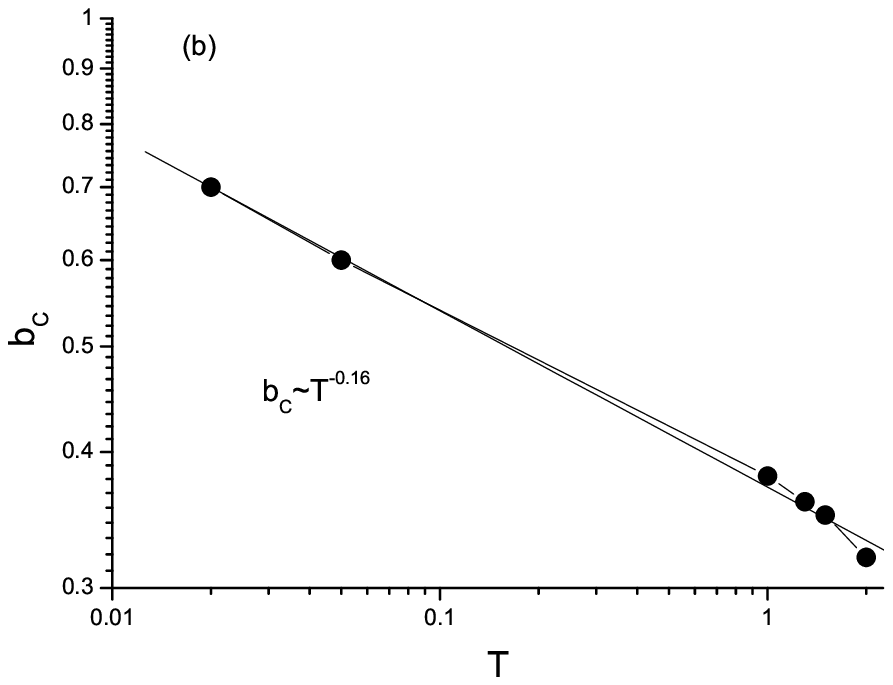}
  \includegraphics[width=8cm,height=6cm]{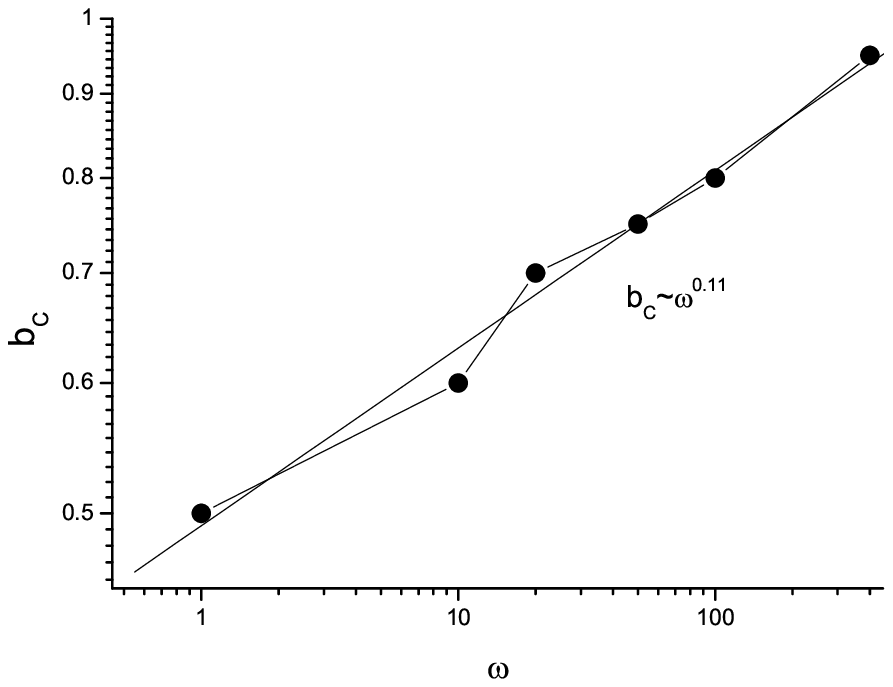}
  \caption{(a)The mean velocity $v$ as a function of parameter $b$ at $a=0.5$, $b_{0}=0.3$, $\Delta=-1.0$, $\omega=10.0$, $T=0.05$, and $L=1.0$.
  $b_{C}$ denotes the optimized value of $b$ for the peak. (b)Dependence of $b_{C}$ on temperature $T$ at $a=0.5$, $b_{0}=0.3$, $\Delta=-1.0$, $\omega=10.0$, and $L=1.0$.
  (b)Dependence of $b_{C}$ on the driving frequency $\omega$ at $a=0.5$, $b_{0}=0.3$, $\Delta=-1.0$, $T=0.05$, and $L=1.0$. The solid lines are the fitting lines.}
  \label{1}
\end{center}
\end{figure}

Figure 6 (a) shows the mean velocity $v$ as a function of the
controlling parameter $b$. The parameter $b$ determines the radius
at the bottleneck. The tube may be close for too small values of
$b$. If the radius at the bottleneck is small few particles can pass
from one cell to another, so the mean velocity is small. When the
bottleneck has infinite radius ($b\rightarrow \infty$), the tube
reduces to a straight one and the effect of tube shape disappears,
so the current goes to zero. Therefore, there exists a value of
$b_{C}$ at which the mean velocity takes its maximum. We also study
how the temperature $T$ and the vibrating frequency $\omega$ affect
the optimized value $b_{C}$.  From the numerical simulations, we
also find the approximate scaling results for $b_{C}\propto
T^{-0.16} \omega^{0.11}$.  It is found that $b_{C}$ decreases with
temperature $T$ and increases wit the vibrating frequency $\omega$.

\begin{figure}[htbp]
  \begin{center}\includegraphics[width=8cm,height=8cm]{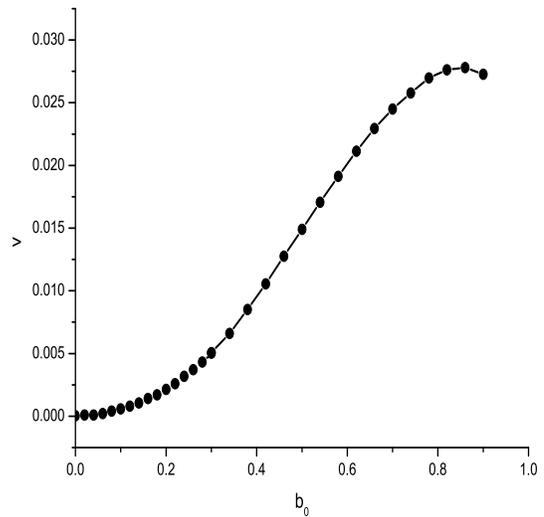}
  \caption{The mean velocity $v$ as a function of vibrating amplitude $b_{0}$ at $a=0.5$,$b=0.9$, $\Delta=-1.0$, $\omega=10.0$, $T=0.05$, and $L=1.0$. }
  \label{1}
\end{center}
\end{figure}

\indent Figure 7 gives the mean velocity $v$ as a function of the
vibrating amplitude $b_{0}$. When $b_{0}\rightarrow 0$, the
particles can not experience the wall vibrating, so the current goes
to zero. On increasing the vibrating amplitude $b_{0}$, the mean
velocity increases. However, for large values of $b_{0}$, the tube
may be close at some time, then the current decreases. We must point
out that for very large value of $b_{0}$, the profiles of tube wall
may be exchanged, this is not true for a real system.

\section{Concluding Remarks}
\indent In this paper, we study the transport of overdamped Brownian
particles moving in an asymmetrically periodic tube with the wall
vibration. From the numerical simulation we can find that the
transverse wall vibration can induce a net longitudinal current
 in an asymmetrically periodic tube. We can have current reversals by changing the sign
of $\Delta$, the asymmetry of the tube. The sign of mean velocity
for low frequencies is opposite to that for high frequencies.
Therefore, we can change the direction of the current by suitably
tailoring the vibrating frequency. Thus, varying the asymmetry of
the tube and the vibrating frequency are the two ways of inducing
current reversals. These peculiar reversals are due to different
strength and asymmetry of relaxation processes. The symmetry of
relaxation processes changes with the system parameters, so the
current may changes its direction when the parameters are changed.
In addition, there exists an optimized value of temperature at which
the mean velocity  takes its maximum value, which indicates that the
thermal noise may facilitate the particles transport.

\indent Clearly, the model is too simple to provide a realistic
description of real systems, however, the results we have presented
have a wide application in may processes, such as diffusion of ions
and macromolecular solutes through the channels in biological
membranes\cite{c1}, transport in zeolites\cite{c2}and nanostructures
of complex geometry \cite{c3}, controlled drug release\cite{c4}, and
diffusion in man-made periodic porous materials\cite{a12}. It is
very important to understand the novel properties of these confined
geometries, zeolites, biological channels, nanoporous materials, and
microfluidic devices, as well as the transport behavior of species
in these systems.

\indent The author is very grateful to the two anonymous reviewers
for the valuable comments and suggestions. This work was supported
in part by National Natural Science Foundation of China with Grant
No. 30600122 and GuangDong Provincial Natural Science Foundation
with Grant No. 06025073.

\end{document}